Invariance under Structure Translation as the Origin of Host Immune Capacity Conservation from Noether's Theorem

*Yexing Chen [1]* and Peng Cao [1,2]**

[1] State Key Laboratory of Technologies for Chinese Medicine Pharmaceutical Process Control and Intelligent Manufacture, Nanjing University of Chinese Medicine, Nanjing, 210000, PR China.

[2] Jiangsu Key Laboratory for Pharmacology and Safety Research of Chinese Materia Media, Nanjing University of Chinese Medicine, Nanjing, 210000, PR China

[*] Corresponding author. Email: Xycyx@njucm.edu.cn, Cao_peng@njucm.edu.cn

**Abstract**

Biological processes are reproducible, including immunity. Different healthy individuals can exhibit similar disease courses and protective serum antibody levels when confronting the same pathogen. This widely observed phenomenon is the foundation of almost all immunological experiments and clinical trials. This is regarded as a property of immune homeostasis. But why? Modern studies reveal that the core processes in adaptive immunity are governed predominantly by stochastic events, including V(D)J recombination and affinity maturation. The final composition of the serum antibody repertoire in each individual is also unique. What constrains these microscopic random processes to converge onto consistent, predictable macroscopic trajectories? What is the consistency behind this degeneracy? Here, we address these questions by introducing a theoretical framework based on Lagrangian analytical mechanics. We reformulate adaptive immune recognition as a dynamical system evolving in a high-dimensional immunological state space, with generalized coordinates representing the immune repertoire configuration and the effective antigenic structure. We identify a continuous symmetry: the system's action remains invariant under specific translations within the antigenic structure space. Noether's theorem dictates a consistency constraint, $I$, which acts as a physical constraint on the system's evolution. It mandates that, irrespective of the specific stochastic trajectory taken at the molecular level, the system must converge to the same terminal state for a given challenge. This framework yields a falsifiable prediction: if two organisms from different species are configured with identical consistency constraint $I$ and its associated parameters, after appropriate normalization for allometric scaling of cellular metabolic rates, their normalized immune response trajectories must be statistically indistinguishable. This unified principle offers a first-principles explanation for immune memory, tolerance, immunodominance, and immune imprinting. We transform the concept of immune homeostasis from a phenomenological description into a quantifiable, consistent physical entity called *Immune Capacity*, and establish a foundational framework for a unified, predictive immunology.

## 1. Introduction

Different healthy individuals often display comparable disease progression and protective serum antibody levels upon exposure to the same novel pathogen.[1, 2] This widely observed phenomenon is the foundation of almost all immunological experiments, vaccine trials, and clinical practices. Its reproducibility is frequently attributed to the shared state of immune homeostasis among these individual, yet this explanation remains phenomenological rather than mechanistic.

But why are adaptive immune responses reproducible? Modern immunology has revealed a profound paradox. At the molecular and cellular levels, the processes that generate adaptive immune responses are stochastic. The primary repertoire of B-cell and T-cell receptors is generated through V(D)J recombination, a random process of combinatorial gene segment assembly and junctional nucleotide insertion and deletion.[3] Affinity maturation in germinal centers proceeds via somatic hypermutation where the trajectory of any given B-cell clone is unpredictable. Consequently, the final composition of the serum antibody repertoire in each individual is unique.[4, 5] High-throughput sequencing studies have shown that different individuals immunized with the same antigen produce distinct immunoglobulin repertoires, yet they converge on the same functional outcome of protection.[6, 7] The question thus arises: what physical principles constrain these microscopic random processes to converge onto consistent, predictable macroscopic trajectories? What is the hidden consistency behind this apparent degeneracy? The reproducibility of immune outcomes across individuals is a regularity that demands explanation.

This degeneracy is a fundamental source of robustness and evolvability in complex biological systems. It is a well-characterized feature of immune system: multiple T-cell receptors can recognize the same peptide-MHC complex,[8] multiple B-cell lineages can generate antibodies targeting the same epitope, and different individuals can mount protective responses through entirely different cell compositions. However, while degeneracy describes the phenomenon, what enforces the constraint that different paths lead to the same destination?

Existing theoretical frameworks offer partial but incomplete answers. The diversity of the immune repertoire and the time scales of immune responses can be described by statistics. Entropy quantifies the number of accessible microstates and can predict the statistical tendency of a system toward equilibrium distributions. Network models capture the signaling and regulatory interactions among immune cells, revealing how positive and negative feedback loops generate stable states. Markov chain descriptions of clonal selection and somatic hypermutation provide probabilistic frameworks for affinity maturation, demonstrating how mutation-selection cycles drive the population toward higher-affinity states. But these frameworks do not answer the fundamental question: why must the system evolve in this way? They describe how the system behaves but do not provide a first-principles constraint that explains why the system is obliged to converge to the same macroscopic end state from different microscopic starting conditions. A global, parameter-independent constraint on the system's evolution is missing, which bounds the ensemble of possible paths irrespective of the specific stochastic trajectory taken. Such a principle would answer not only how the system evolves, but why its macroscopic behavior is so robustly reproducible? We propose that this principle is a continuous symmetry and its associated consistency constraint, derived from the Lagrangian formalism of analytical mechanics and Noether's Theorem.

Each individual’s response to a given pathogen constitutes one realized path, including processes from initial antigen encounter to clonal expansion, affinity maturation, and resolution. A historically successful response that clears the infection and establishes long-lived memory represents a particular solution to the equations of motion, but it is by no means the only mathematically allowed trajectory. The ensemble of all conceivable paths, weighted by their respective actions, defines a path functional. Within this functional landscape, the empirically observed convergence of diverse microscopic trajectories toward the same macroscopic outcome implies that the path functional must possess a constraint irrespective of initial microscopic details. Noether’s theorem, when applied to the symmetry of this path functional, provides exactly such a constraint: it identifies an invariant quantity that all physically realizable paths must respect, thereby forcing the ensemble to cluster around a unique macroscopic solution for a given challenge. In this view, the successful clearance trajectory is not an accident of initial conditions but a manifestation of a deeper conservation law that governs the entire immune dynamics.

We start from the adaptive immune recognition and construct a Lagrangian formulation, treating the immune system as a dynamical system evolving in a high-dimensional immunological state space with generalized coordinates representing the immune repertoire configuration and the effective antigenic structure. We identify a continuous symmetry: the system's action remains invariant under specific translations within the antigenic structure space. By Noether's theorem, these symmetries dictates consistency constraints, which we together term the *Immune Capacity* I. This *Immune Capacity* acts as a physical constraint on the system's evolution, mandating that the system must converge to the same macroscopic trajectories for a given challenge irrespective of the stochastic molecular randomness.

The framework unifies diverse immunological phenomena , including immune memory, tolerance, immunodominance, and immune imprinting. By transforming the strength of immune homeostasis from a phenomenological description into a quantifiable, consistent physical entity, we establish a foundational framework for a unified, predictive immunology grounded in first principles. The framework yields a falsifiable prediction:  If two subjects from different species are configured to possess identical systemic parameters, then when challenging with stimulation having identical immunogenicity, their subsequent immune responses must converge to statistically indistinguishable trajectories after normalization by cellular turnover rates or metabolic scaling.

## 2. Analytical Mechanics of Immune Recognition

To establish a dynamical framework for adaptive immunity, we must first define the space in which the immune system evolves and identify the degrees of freedom that capture its macroscopic behavior. We start from the molecular recognition event between a single antigen and a single receptor, to the ensemble behavior of a diverse lymphocyte population, and finally to the high-dimensional state space that describes the system at the organismal level. This layered construction ensures that each abstraction is rooted in experimentally established biophysical principles and provides a clear link back to the central puzzle of this work—how microscopic randomness gives rise to reproducible macroscopic outcomes.

### 2.1. Multi-scale Potential Energy Surface Theory

The immune response is initiated at the molecular level by the specific binding between an antigenic epitope and a cognate receptor (e.g., Toll Like receptor (TLR), B-cell receptor (BCR) or T-cell receptor (TCR)). This physicochemical process is conceptualized within a potential energy curve.[9, 10] In the unbound state, both the antigen and receptor molecules reside in their respective low-energy steady states. Successful binding requires the system to overcome an activation energy barrier, a process driven by thermal fluctuations or cellular motions, leading to a conformational change and a transition to a more stable, lower-energy bound state (**Figure**. 1 A-B). The height of this barrier governs the binding probability, whereas the energy difference ($\Delta E$) between the initial and final states is determined jointly by the intrinsic structural potential of the antigen and the receptor, as well as the coupling potential released upon their interaction as the system relaxes into a lower-energy state. This $\Delta E$ represents the driving force that propels downstream biochemical cascades, ultimately culminating in cellular activation, proliferation, and differentiation.

At the cellular population level, the adaptive immune system maintains a vast repertoire of receptors through stochastic genetic recombination. Consequently, the response to a specific antigen is not mediated by a single receptor type but by an ensemble of clones with a continuous distribution of binding affinities.[11] This results in a corresponding distribution of activation energy barriers across the responsive cell population, thereby extending the potential energy curve into a potential energy surface (PES). Upon initiation of the response, the system spontaneously evolves toward potential wells in the PES, following the principle of least action, with multiple potential wells corresponding to multiple possible reaction pathways (**Figure**. 1 B-C). The statistical properties of this distribution, including its mean, variance, and shape, collectively determine the population's overall response intensity and sensitivity.

At the macroscopic scale, the immune system must be prepared to respond to a near infinite universe of potential antigens. It employs a three-pronged strategy: deploying a consistent innate immune response to handle low-level stimuli and mitigate the intensity of stronger challenges,[12-15] generating pattern-recognition receptors (e.g., TLRs) with pre-defined, low-energy barriers for conserved pathogenic structures,[3, 13, 16] while also maintaining an immense diversity of somatically rearranged receptors (e.g., TCRs and BCRs) to cover unforeseen threats.[16, 17] This creates a high-dimensional recognition space where each potential antigen maps to a specific PES, activating a unique subset of the total repertoire. Thus, the interaction between the antigenic universe and the immune system is no longer described by individual energy curves but emerges as a high-dimensional PES, which encodes the system's entire landscape of reactivity.

The PES describes the possibility space of immune responses; what remains to be explained is why the system's dynamics on this landscape are reproducible. This is the question to which the symmetry-based conservation law will provide an answer.

## 2.2 The Generalized Coordinates and the State Space

To formulate the immune system as a dynamical system amenable to Lagrangian mechanics, we must first specify a set of generalized coordinates that uniquely and completely describe its instantaneous configuration. We propose a total generalized coordinate vector partitioned into three physically and biologically distinct classes:

$$\boldsymbol{q} = (q_{\mathrm{ab}}, q_{\mathrm{ag}}, r, \xi)$$

$q_{\mathrm{ab}} = (q_{\mathrm{ab1}}, q_{\mathrm{ab2}}, \ldots q_{\mathrm{abN}})$ are receptor conformation coordinates, including the three-dimensional folding pattern of each amino acid, and the binding spatial angles when bound to the antigen, all of which can be determined by structural biology methods such as X-ray diffraction and cryo-electron microscopy.

$q_{\mathrm{ag}} = (q_{\mathrm{ag1}}, q_{\mathrm{ag2}}, \ldots q_{\mathrm{agN}})$ are antigen conformation coordinates, including the spatial conformation of the antigen, determined by the same methods as for the receptor.

$r = (r_1, r_2 \ldots, r_{\mathrm{N}})$ are spatial coordinates; they describe the physical distance and relative orientation between the antibody and the antigen. These coordinates determine whether the two molecules are close enough to interact. Binding cannot occur when the molecules are far apart; as they approach, the interaction potential becomes significant. These are also determined by structural biology methods.

$\xi = (\xi_1, \xi_2 \ldots, \xi_{\mathrm{N}})$ describes the macroscopic state of the immune cell population. The index $\boldsymbol{N}$ runs over all possible clonotypes present in the repertoire. $\xi_k$ represent the population of cells whose $q_{ab}$ are the same, and can be measured by biological methods such as cell counting and gene sequencing.

The external parameter $N_{ab}$ denotes the number of antigen molecules.

The derivatives of these coordinates represent the rate of conformational change of the antibody and the antigen epitope; the relative molecular speed; and the rate of clonal expansion or contraction, respectively.

$$\dot{q}_{\mathrm{Ab}} = \frac{\mathrm{d}q_{\mathrm{Ab}}}{\mathrm{d}t}, \dot{q}_{\mathrm{Ag}} = \frac{\mathrm{d}q_{\mathrm{Ag}}}{\mathrm{d}t}, \dot{r} = \frac{\mathrm{d}r}{\mathrm{d}t}, \dot{\xi} = \frac{\mathrm{d}\xi}{\mathrm{d}t}$$

The entire set $\boldsymbol{q}$ thus spans a high-dimensional state space in which the immune system evolves. Each point in this space corresponds to a complete specification of the system's instantaneous configuration: every antibody shape, every antigen conformation, every spatial arrangement, and every clone size. The dynamics of the system is the trajectory through this space over time.

## 2.3 Generalized Masses and the Kinetic Energy

In our generalized **coor**dinate framework, each generalized coordinate is associated with its own generalized mass, which measures how difficult it is to change the corresponding coordinate's velocity.

Mass associated with antibody conformation $\boldsymbol{M}_{\mathrm{ab}}$ **:** The resistance to conformational changes in protein shape is measured by the moment of inertia of the relevant atomic groups. For a collection of atoms, the total moment of inertia is the sum over all atoms. Larger, bulkier amino acid side chains like tryptophan contribute larger inertial resistance to conformational change, whereas smaller side chains like glycine contribute very little. These can be determined by structural biology methods.

Mass associated with antigen conformation $\boldsymbol{M}_{\mathrm{ag}}$ **:** It measures the resistance of the antigen's epitope to conformational change. This is dominated by the hinge regions of multidomain proteins, for example the RBD of the SARS-CoV-2 spike protein. These can also be determined by structural biology methods.

Mass associated with spatial coordinates $\boldsymbol{M}_{\mathrm{r}}$ **:** The spatial motion of two molecules approaching each other is governed by their actual physical masses. These can be measured by determining the protein mass.

Mass associated with cellular abundance $\boldsymbol{M}_{\xi}$ : It measures the difficulty for immune cells to activate and proliferate. This is related to the mechanical force required for signal transduction contacts, the intracellular chemical concentrations, and the duration of the cell cycle for expansion.

With these definitions, the total kinetic energy of the immune system is:

$$T = \frac{1}{2}\sum_{k\in A} \xi_k \, M_{\mathrm{ab}} \dot{\mathbf{q}}_{\mathbf{Ab}}^{\mathbf{2}} + \frac{1}{2} N_{ab} M_{\mathrm{ag}} \dot{\mathbf{q}}_{\mathbf{Ag}}^{\mathbf{2}} + \frac{1}{2} M_r \dot{\boldsymbol{r}}^{\mathbf{2}} + \frac{1}{2} M_{\xi} \dot{\xi}^2$$

where the squared velocities are shorthand for the squared Euclidean norms of the corresponding vectors.

## 2.4 The Coupling Function $G$ and the Active Subset

Not every antibody can bind every antigen. An antibody specific for the influenza hemagglutinin will not recognize the SARS-CoV-2 spike protein. Among a population of B cells, only those whose receptors happen to match the incoming antigen will be activated. This selectivity is the central feature of adaptive immunity. We capture this selectivity through a mathematical object we call the coupling function $G$.

Consider a single antibody clone specified by its conformation coordinates $\boldsymbol{q}_{Ab}$. Its binding interface is defined by its three-dimensional arrangement. When confronted with an antigen whose conformation is specified by $\boldsymbol{q}_{Ag}$, the strength and specificity of their interaction are determined by how well the two surfaces fit together, the shape complementarity, the electrostatic match, the hydrogen-bonding potential, and the hydrophobic contacts. All of these molecular determinants are aggregated into a single scalar function:

$$G = G(\boldsymbol{q}_{Ab}, \boldsymbol{q}_{Ag})$$

This function is the mathematical measure of immunogenicity, the ability of a given antigen-antibody pair to generate an effective immune signal. It mediates the intrinsic structural recognition between the antibody paratope and the antigen epitope, and is independent of their spatial separation.

Among the vast repertoire of lymphocyte clones, only those whose receptors can engage the antigen with non-negligible strength will participate in the response. The remainder are unaffected by the antigen's presence. This selective recruitment is captured by defining the active subset $\mathcal{A}$, a clone $k$ is active only if its coupling function with the current antigen is non-zero.

$$\mathcal{A} = \{k | G_k(\boldsymbol{q}_{Ab}, \boldsymbol{q}_{Ag}) \neq 0, \xi_k > 0, N_{ab} > 0\}$$

The number of active clones $\mathcal{A}$depends on the antigen's conformation: different antigens $\boldsymbol{q}_{Ag}$ will activate different subsets of the repertoire. This formalizes the concept of antigen specificity. Second, if the antigen $\boldsymbol{q}_{Ag}$ mutates, previously active clones may become inactive and previously inactive clones may become active, this is the mathematical description of antigenic drift.

The interaction potential for receptor and antigen is a function of the structural match $G$, the spatial distance $r$, and the cell abundancy $\xi$ :

$$V = \sum_{k \in A} \xi_k \, \mathcal{U}(G_k, r_k)$$

The function $U$ can be thought of as a typical intermolecular potential well. The exact shape of $U$ does not affect our main theoretical conclusions.

## 2.5 The Complete Potential Energy and Total Lagrangian

The total potential energy $V$ of the immune system is:

$$V = \sum_{k \in A} \xi_k \, V_{self}(\boldsymbol{q}_{\mathrm{ab}}) + N_{ab} V_{intrinsic}(\boldsymbol{q}_{\mathrm{ag}}) + \sum_{k \in A} \xi_k \, \mathcal{U}(G_k, r_k)$$

$V_{self}(\boldsymbol{q}_{\mathrm{ab}}), V_{intrinsic}(\boldsymbol{q}_{\mathrm{ag}})$ are the potential energy when the antibody and antigen adopts its most stable, unbound conformation, and increases as the structure is distorted.

the Lagrangian $L$ is defined as the difference between the kinetic energy $T$ and the total potential energy $V$:

$$L = T - V$$

$$L(\boldsymbol{q}, \dot{\boldsymbol{q}}) = \frac{1}{2} \sum_{k \in A} \xi_k \, M_{\mathrm{ab}} \dot{q}_{\mathbf{Ab}}^{\mathbf{2}} + \frac{1}{2} N_{ab} M_{\mathrm{ag}} \dot{q}_{\mathbf{Ag}}^{\mathbf{2}} + \frac{1}{2} M_r \dot{r}^{\mathbf{2}} + \frac{1}{2} M_\xi \dot{\xi}^2$$
$$- \sum_{k \in A} \xi_k \, V_{self}(q_{\mathrm{ab}}) - N_{ab} V_{intrinsic}(q_{\mathrm{ag}}) - \sum_{k \in A} \xi_k \, \mathcal{U}(G_k, r_k)$$

## 2.6 The Open System: Driving Forces, Dissipative Forces, and Immune States

The immune system is an open system, it constantly exchanges energy and matter with its environment. Energy enters the system in the form of nutrients, oxygen, and signaling molecules (cytokines, hormones). Energy leaves the system through metabolic heat, the removal of dead cells. Matter enters through new lymphocyte production in the bone marrow and thymus; matter leaves through apoptosis and migration into peripheral tissues.

To capture this openness, we include two classes of forces: generalized driving forces and generalized dissipative forces. Generalized driving forces $Q^{in}$ are from cytokines that promote proliferation, growth factors that enhance survival, co-stimulatory signals from helper T cells or innate immune cells, and the presence of the antigen. Generalized dissipative forces $Q^{\mathrm{out}}$ are from cell apoptosis.

The actual behavior of the immune system at any moment is determined by the net generalized force acting on each coordinate:

$$F_{net,i} = Q_i^{in} - Q_i^{\mathrm{out}} - \frac{\partial V}{\partial \mathrm{q}_i}$$

The complete equation of motion with both driving and dissipative forces is:

$$\frac{d}{dt}(\frac{\partial L}{\partial \dot{q}_i}) - \frac{\partial L}{\partial q_i} = Q_i^{\text{net}}$$

This equation is the master equation of our immunological dynamics, from which all the phenomena described above emerge as different solutions corresponding to different balances of the forces, including resting, activation, resolution, memory, hyperactivation, exhaustion, and tolerance.

### 2.7 A Potential Explanation for Existing Immunological Phenomena

With the equations of motion established, the full spectrum of immune phenomena can be interpreted as different dynamical states of the system. Each phenomenon corresponds to a specific balance of the potential forces, driving forces, and dissipative forces.

Homeostasis. In the resting state, the baseline is exactly balanced. All accelerations are zero. The system resides at the minimum of its effective potential for the given baseline input. This is the physical definition of immune homeostasis.

Activation (**Figure**. 1 C). When a acute new pathogen enters, its antigen creates non-zero coupling $G$ for specific clones, making them the active subset $\mathcal{A}$. The binding energy term $V = \sum_{k\in A} \xi_k \mathcal{U}(G_k, r_k)$ creates a driving force: $Q_{antigen}^{in} =- \partial V/\partial q > 0$. The net force $Q^{\text{net}}$ becomes positive, driving cell activation and clonal expansion. This is the activated state of an ongoing immune response.

Affinity Maturation. Somatic hypermutation and selection are stochastic search processes in the antibody conformation coordinate $\boldsymbol{q}_{Ab}$: each mutation is a random jump in conformation space. Then select the antibodies which pulls the conformation toward configurations with deeper potential well.

Vaccination. Vaccination is an artificial perturbation of the potential energy surface. A vaccine introduces a modified antigen $\boldsymbol{q}_{ag}^{vaccine}$ that is immunogenic but non-pathogenic. It activates the active subset $\mathcal{A}$ just as an infection would, carving a potential well to the vaccine antigen and stored through memory. When the actual pathogen later enters, the pre-existing memory well alters the system's trajectory: the principle of least action draws the response toward the pre-optimized path, yielding faster and stronger protection. Vaccination is thus a deliberate engineering of the potential landscape to minimize future pathogenic action.

Memory. After infection, memory cells persist with an easier activation threshold, which means a deeper $V$and smaller mass $M$. Upon re-encounter with the same antigen, it produces a larger acceleration. This is the mechanistic basis of faster and more robust recall responses.

Tolerance and Autoimmunity (**Figure**. 1 D). The host potential $V_{self}$ is shaped such that for self-antigens, the coupling function $G$ remains zero and the active subset $\mathcal{A}$ empty for all self-reactive conformations. This is tolerance. If the potential landscape is altered so that $G$ becomes non-zero, the system is driven toward self-reactive configurations, producing autoimmunity.

Exhaustion. In chronic infections, persistent stimulation progressively increases the difficulty to active clones through upregulation of inhibitory receptors and metabolic insufficiency. With $M$ increasing over time, the clone becomes increasingly difficult to sustain effective, resulting in T-cell exhaustion.

Immunodominance. The potential energy surface $V$ is not homogeneous: some epitopes have deeper and broader potential wells than others. The system preferentially follows paths of least action into the deepest accessible wells. These correspond to immunodominant epitopes. The phenomenon emerges from the topography of the potential landscape.

Immune Imprinting). Upon first infection, the system carves a deep potential energy well in a specific region. When a drifted antigen arrives, the system's trajectory is constrained by this pre-existing well. The principle of least action dictates that the system follows paths near previously optimized trajectories, even if those paths are now suboptimal.

Antibody-Dependent Enhancement (ADE). The antibody may bind with high affinity to a non-neutralizing epitope, producing a deep binding energy well. However, if this binding does not occlude the viral entry machinery, the energy well for viral fusion with the host cell remains intact, The antibody, far from blocking infection, acts as a shuttle that delivers the virus directly to cells bearing Fc receptors. This is the dynamical origin of ADE: a deep binding well coexisting with an unaltered entry well.

In summary, all of these phenomena are different dynamical states arising from the same framework. This unification demonstrates the explanatory power of the analytical mechanics framework. In the following section, we show that this framework additionally yields a consistency constraint via Noether's theorem, providing the first-principles constraint that explains the reproducibility of immune outcomes.

## 3. Evidence for Degeneracy and the Need for a Physical Principle

The section established a dynamical framework capable of describing immune phenomena as trajectories on a potential energy surface. We now turn to the empirical foundation that motivates the core theoretical proposal of this work: the hidden symmetry and its associated conservation law underlie the reproducibility of immune responses.

### 3.1. Multi-scale Evidence for Degeneracy

Degeneracy is a well-characterized feature of the immune system. In the immune context, degeneracy manifests at multiple observational scales, each providing evidence that different microscopic configurations converge to the same macroscopic outcome.

Molecular scale. A single T-cell receptor can recognize multiple distinct peptide-MHC complexes; a given peptide-MHC complex can engage diverse T-cell receptors. Structural studies have shown that T-cell recognition is characterized by substantial conformational degeneracy. Innate immune receptors exhibit pattern-directed rather than sequence-specific recognition: cGAS detects the phosphate-deoxyribose backbone of DNA irrespective of nucleotide sequence; TLR4 recognizes the conserved lipid A moiety of lipopolysaccharide across diverse Gram-negative bacteria; and the NLRP3 inflammasome is triggered by a broad set of structurally distinct 'danger signals' that share the ability to perturb ion homeostasis. In all these cases, different molecular structures converge on the same functional outcome.

Clonal scale. Inbred mice of the same strain, housed in identical conditions, generate non-identical B-cell and T-cell receptor repertoires in response to the same immunization, yet show no significant difference in overall protection. High-throughput repertoire sequencing has revealed that different individuals immunized with the same antigen produce immunoglobulin repertoires that populate distinct branches of phylogenetic trees. The specific V(D)J recombination events, the exact somatic hypermutations, and the resulting receptor sequences are almost entirely non-overlapping between individuals. Yet these distinct repertoires converge on the same protective antibody titers that neutralize the pathogen. The protection is determined not by the specific clonal identities but by the statistical equivalence of the overall receptor distribution at the population level.

Individual and population scale. At the organismal level, the reproducibility of immune outcomes is the foundation of clinical practice and experimental design. Patients infected with the same respiratory pathogen like influenza, respiratory syncytial virus, rhinovirus often present with nearly identical symptom profiles (fever, cough, malaise) and time course despite having entirely different genetic backgrounds, different prior exposure histories, and different specific antibody repertoires. Routine clinical diagnostics can identify a viral category but cannot specify the exact pathogen strain, unless a molecular-level diagnosis is performed, because the host response patterns are conserved.

Temporal scale: When infected with the same strain of influenza virus, the healthy 20-year-old and healthy 22-year-old population will exhibit nearly identical disease courses statics: the incubation period, the symptomatic peak, the resolution, and the trajectory of antibody titer rise to protective levels. The kinetics are indistinguishable when plotted against days post-immunization. The calendar age difference is irrelevant to the dynamics, provided the baseline immune homeostasis are comparable. This suggests that the equations governing the immune response do not depend on absolute chronological time

These observations at molecular, clonal, individual, and temporal scales point to a universal pattern: immune outcomes are reproducible not because the microscopic details are reproducible, but because some higher-order constraint forces different microscopic trajectories to converge to the same macroscopic endpoint. This is the essence of degeneracy. What imposes this constraint?

### 3.2 What The Existing Frameworks Describe

Several powerful theoretical frameworks have been applied to the immune system. Each captures important aspects of its behavior. But none provides a first-principles explanation for why different microscopic paths must lead to the same macroscopic outcome. They describe ‘how’ the system behaves, but not ‘why’ its behavior is so robustly reproducible.

Statistical mechanics and entropy has been extensively applied to characterize the immune repertoire. The Shannon entropy of a BCR or TCR repertoire quantifies its diversity: a highly diverse repertoire has high entropy; a clonally expanded repertoire has low entropy. However, entropy is a state function, it describes the number of accessible microstates for a given macrostate, but it does not constrain which microstate the system actually occupies, nor does it guarantee convergence from different initial conditions. Two entirely different stochastic trajectories can yield the same entropy value; entropy does not ensure that they must converge to the same functional outcome. The observation that twins with distinct antibody repertoires (different sequences, different clonal compositions) can have the same Shannon entropy is expected, but entropy does not explain why they both must produce the same protective titer. It answers the question "what is the most probable state?" but not "why must the system arrive at that state regardless of where it started?"

Systems biology has produced large-scale, integrative models of the immune system that connect molecular interactions to cellular phenotypes and organism-level behaviors.[18] These models reconstruct regulatory and signaling networks from high-throughput data and simulate their dynamics using ordinary differential equations, Boolean networks, or logic-based formalisms. Network theory has revealed that these immune networks possess characteristic topological features like scale-free degree distributions and modular organization that shape how signals propagate through the system. These models have demonstrated that immune networks can exhibit stable configurations of gene expression, signaling activity, or cell population sizes toward which the system converges regardless of initial conditions. For example, models of T-cell differentiation have shown that the regulatory network governing Th1/Th2 polarization can flow into one or the other depending on the cytokine environment. The limitation of this approach is that they are model-dependent constructs. They emerge from specific parameterizations of specific differential equations and are inferred from specific experimental conditions. They do not explain why the biological system must evlove nor does it explain why it is robust across individuals whose biochemical parameters vary substantially. Network models describe how the system can converge, but they do not derive the necessity of convergence from a first-principles invariant.

The stochastic nature of V(D)J recombination, somatic hypermutation, and clonal selection has been formalized using stochastic models like Markov chain. These models describe the probability of transitioning from one state to another. Markov models have successfully reproduced power-law distributions of repertoire diversity and the kinetics of affinity maturation. But the transition probabilities in these models are phenomenological inputs. They are inferred from experimental data rather than derived from a deeper principle. The models describe what happens but they do not explain why the transition probabilities take the values they do, nor why the system is constrained to converge to the same steady state across different individuals who have different starting repertoires and different mutation histories. The Markov chain is a description of the stochastic process, not an explanation of its invariance.

All of these frameworks describe how the immune system behaves but do not explains why the system behave in this way. They describe the mechanisms of convergence but do not explain the origin of convergence. There lacks a physical principle that acts as a global constraint on the system's evolution, a principle that guarantees the reproducible of macroscopic outcome. Such a principle would answer not only how the system evolves, but why its macroscopic behavior is so robustly reproducible.

### 3.3 From Degeneracy to Symmetry

The fact that different molecular structures, different clonal compositions, different individuals, and different ages all converge to the same immune outcomes strongly suggests that this convergence is not coincidental. It is too robust, too universal, to be accidental. It must be governed by a physical principle.

We propose that this principle is a symmetry, an invariance of the system's dynamics under certain transformations. We hypothesize that the immune system exhibits a symmetry in the space of immunological configurations. There exists a continuous family of transformations that leave the system's action $S = \int L \, dt$ invariant. These transformations may shift the antibody conformation $\boldsymbol{q}_{ab}$ while simultaneously adjusting the antigen conformation $\boldsymbol{q}_{ag}$, the spatial arrangement $\mathbf{r}$, and the cellular abundances $\boldsymbol{\xi}$ in a coordinated manner, such that the Lagrangian changes by at most a total time derivative.

This symmetry is the mathematical formalization of degeneracy: different points in the full state space are dynamically equivalent. This is precisely the property we need to explain why different molecular structures, different clonal compositions, and different absolute ages all produce the same immune outcomes. They are all connected by transformations that leave the dynamics invariant.

Then Noether's theorem guarantees the existence of a consistency constraint associated with each continuous symmetry direction. These consistency constraints, which we will collectively refer to as the *Immune Capacity* $\mathbf{I}$, act as global constraints on the system's evolution. They are the physical embodiment of the reproducibility: quantities that must remain constant throughout the instances, irrespective of the specific stochastic trajectory followed at the microscopic level. The existence of such invariants is precisely the missing principle that bounds the ensemble of possible paths and forces convergence to a reproducible macroscopic outcome.

## 4. From Degeneracy to Symmetry: The Consistentency of Immune Response

We posit that the observed consistency of immunity is not coincidental but must be governed by a fundamental physical principle. We now demonstrate how this symmetry mandates the existence of consistency constraints via Noether's theorem, which we identify as the physical embodiment of *Immune Capacity*.

### 4.1. Spatial Continuous Symmetry

Remarkably, the existence of a structural symmetry can be derived purely from the structure of the Lagrangian, without relying on any biological observation beyond the definitions of the coordinates and potentials.

We begin with the simplest case: consider a transformation that acts only on the antibody conformation coordinates $\mathbf{q}_{ab}$, shifting each active clone's conformation along a direction $\boldsymbol{\eta}_{ab}$that leaves the coupling function $G_k = G(\mathbf{q}_{ab}, \mathbf{q}_{ag})$ invariant. If the antibody structural "looseness" $V_{self}(\mathbf{q}_{ab})$ remains unchanged under this shift, and the antigen conformation, spatial coordinates, and clone sizes are held fixed, then the total Lagrangian $L$ is invariant: . Explicitly, the transformation is:

$$\mathbf{q}_{ab} \to \mathbf{q}_{ab} + \epsilon \boldsymbol{\eta}_{ab}, \delta\mathbf{q}_{ag} = 0, \delta\mathbf{r}_k = 0, \delta\xi_k = 0$$

$$\delta L = 0$$

$$\sum_{k\in\mathcal{A}} \xi_k M_{ab} \dot{\mathbf{q}}_{ab}^{(k)} \cdot \boldsymbol{\eta}_{ab} + \int_0^t \sum_i Q_i^{net}(s)\eta_i ds = \text{Consistent}$$

This is the mathematical expression of a clinically well-known phenomenon: genetically identical twins, who share the same baseline immune architecture (same $V_{self}$, same $G$, same other parameters), can generate entirely different antibody repertoires (different $\mathbf{q}_{ab}$ ) in response to the same infection, yet exhibit identical disease courses. The microscopic paths diverge, but the combination of conformational momentum and integrated force work remains the same for both individuals, because the symmetry is rooted in the shared structure of their Lagrangians.

Now consider a more general transformation that shifts both antibody and antigen conformations simultaneously, again along directions that preserve $G$, while maintaining the intrinsic folding energies $V_{self}$ and $V_{antigen}$ unchanged. The spatial coordinates and clone sizes are held fixed. The transformation is:

$$\mathbf{q}_{ab} \to \mathbf{q}_{ab} + \epsilon\boldsymbol{\eta}_{ab}, \mathbf{q}_{ag} \to \mathbf{q}_{ag} + \epsilon\boldsymbol{\eta}_{ag}, \delta\mathbf{r}_k = 0, \delta\xi_k = 0$$

Again the consistency quantity becomes:

$$\sum_{k\in\mathcal{A}} \xi_k M_{ab}\dot{\mathbf{q}}_{ab}^{(k)} \cdot \boldsymbol{\eta}_{ab} + N_{ab}M_{ag}\dot{\mathbf{q}}_{ag} \cdot \boldsymbol{\eta}_{ag} + \int_0^t \sum_i Q_i^{net}(s)\eta_i ds = \text{Consistent}$$

This corresponds to the clinical observation that different influenza subtypes—H1N1, H3N2, and influenza B—which have distinct antigen conformations $\mathbf{q}_{ag}$ and thus engage different antibody clones, but produce nearly identical symptom profiles in infected patients. The antigen conformation differs, the antibody clones differ, but the sum of conformational momenta for both antibody and antigen, plus the integrated work of the net forces is the same for all these infections, because the symmetry condition $\delta L = 0$ holds across them.

The most general case allows all coordinates to shift simultaneously: antibody conformations, antigen conformations, spatial positions, and clone sizes. The transformation is:

$$\mathbf{q}_{ab}^{(k)} \to \mathbf{q}_{ab}^{(k)} + \epsilon\boldsymbol{\eta}_{ab}, \mathbf{q}_{ag} \to \mathbf{q}_{ag} + \epsilon\boldsymbol{\eta}_{ag}, \mathbf{r}_k \to \mathbf{r}_k + \epsilon\boldsymbol{\eta}_r, \boldsymbol{\xi}_k \to \boldsymbol{\xi}_k + \epsilon\boldsymbol{\eta}_\xi$$

This does not require each potential term to be separately invariant; rather, changes in the self-energies, intrinsic energies, host potentials, and binding energies must compensate each other so that the sum is zero. Under this general symmetry, the consistency quantity is:

$$\delta L = 0$$

$$\begin{aligned}
\delta T = &\frac{1}{2}\sum_{k\in\mathcal{A}} M_{ab}\left[(\delta\xi_k)\dot{\mathbf{q}}_{ab}^{(k)2} + \xi_k \cdot 2\dot{\mathbf{q}}_{ab}^{(k)} \cdot (\delta\dot{\mathbf{q}}_{ab}^{(k)})\right] \\
&+ \frac{1}{2}N_{ab}M_{ag}\left[2\dot{\mathbf{q}}_{ag} \cdot (\delta\dot{\mathbf{q}}_{ag})\right] \\
&+ \frac{1}{2}\sum_{k\in\mathcal{A}} M_r\left[(\delta\xi_k)\dot{\mathbf{r}}_k^2 + \xi_k \cdot 2\dot{\mathbf{r}}_k \cdot (\delta\dot{\mathbf{r}}_k)\right] \\
&+ \frac{1}{2}M_\xi\sum_{k\in\mathcal{A}}\left[2dot\xi_k \cdot (\delta\dot{\xi}_k)\right]
\end{aligned}$$

$$\delta V = \sum_{k\in\mathcal{A}} \left[ \delta\xi_k V_{self} + \xi_k \frac{\partial V_{self}}{\partial \mathbf{q}_{ab}^{(k)}} \cdot \delta \mathbf{q}_{ab}^{(k)} \right]$$

$$+ N_{ab} \frac{\partial V_{intrinsic}}{\partial \mathbf{q}_{ag}} \cdot \delta \mathbf{q}_{ag}$$

$$+ \sum_{k\in\mathcal{A}} \left[ \delta\xi_k \mathcal{U} + \xi_k \frac{\partial \mathcal{U}}{\partial r_k} \delta r_k \right]$$

The full structural consistency quantity is:

$$\sum_{k\in\mathcal{A}} \left[ \xi_k M_{ab} \dot{\mathbf{q}}_{ab}^{(k)} \cdot \boldsymbol{\eta}_{ab} + \xi_k M_r \dot{\mathbf{r}}_k \cdot \boldsymbol{\eta}_r + M_\xi \dot{\xi}_k \eta_\xi \right] + N_{ab} M_{ag} \dot{\mathbf{q}}_{ag} \cdot \boldsymbol{\eta}_{ag}$$

$$-\frac{1}{2} M_{ab} \eta_\xi \dot{\mathbf{q}}_{ab}^{(k)2} - \frac{1}{2} M_r \eta_\xi \dot{\mathbf{r}}_k^2 + \int_0^t \sum_i Q_i^{net}(s) \eta_i ds = \text{Consistent} = I_{\mathbf{spatial}}$$

This is the full structural consistency quantity $I_{Spatial}$. It is not a conserved quantity in the strict sense, its value changes over time as the system exchanges energy and matter with its environment through driving and dissipative forces, as captured by the integral term. However, its evolution is uniquely determined by the macroscopic parameters and is entirely independent of the microscopic stochastic details. Two individuals with the same macroscopic parameters and the same external challenge will have identical $I_{Spatial}$ trajectories, even though their underlying receptor sequences and clonal compositions are entirely distinct. Thus, $I_{Spatial}$ consistency quantity that imposes the same global constraint on every individual sharing the same macroscopic initial conditions. The microscopic paths differ, but the macroscopic constraint is the same for all, providing the first-principles explanation for the reproducibility of immune responses across diverse individuals.

### 4.2. The Uniqueness of Energy Change

Unlike closed conservative systems where time-translation invariance yields energy conservation via Noether's theorem, the immune system is fundamentally open and dissipative, continuously exchanging energy and matter with its environment. This renders strict temporal consistency inapplicable to adaptive immunity.

We now examine cumulative energy invested in the immune response. The Hamiltonian represents the total energy of the system:

$$H = \sum_i p_i \dot{q}_i - L,$$

$$\dot{H} = \sum_i Q_i^{net} \dot{q}_i, H(t) - H(t_0) = \int_{t_0}^t \sum_i Q_i^{net}(s) \dot{q}_i(s) \, ds$$

$$I_{Temporal} = \Delta H = \int_{t_0}^{t} \sum_i Q_i^{net}(s)\, \dot{q}_i(s)\, ds$$

This quantity  is also a consistent quantity: for any two systems with identical initial conditions and identical net forces $Q_i^{net}$ over time, the energy change $\Delta H$ must be identical. The disease course will superimpose when plotted against $\Delta t$, the time elapsed since antigen encounter. This prediction is strongly supported by clinical experience: the characteristic time course of influenza is remarkably consistent across healthy adults of similar baseline immune status, regardless of their exact chronological age. Aging causes a gradual time-dependent drift in the parameter set, leading to the breakdown of symmetry.

$$\mathbf{I} \equiv (I_{\text{Spatial}}, I_{\text{Temporal}})$$

$$I_{\text{Spatial}} = \sum_{k\in\mathcal{A}} \left[\xi_k M_{ab} \dot{\mathbf{q}}_{ab}^{(k)} \cdot \boldsymbol{\eta}_{ab} + \xi_k M_r \dot{\mathbf{r}}_k \cdot \boldsymbol{\eta}_r + M_\xi \dot{\xi}_k \eta_\xi\right] + N_{ab} M_{ag} \dot{\mathbf{q}}_{ag} \cdot \boldsymbol{\eta}_{ag} - \frac{1}{2} M_{ab} \eta_\xi \dot{\mathbf{q}}_{ab}^{(k)2} - \frac{1}{2} M_r \eta_\xi \dot{\mathbf{r}}_k^2 + \int_0^t \sum_i Q_i^{net}(s) \eta_i ds$$

$$I_{\text{Temporal}} = \int_{t_0}^{t} \sum_i Q_i^{net}(s)\, \dot{q}_i(s)\, ds$$

Together, We provide a complete first-principles explanation for the reproducibility of immune responses across individuals: the structural symmetry $I_{\text{Spatial}}$ explains why different microscopic configurations yield the same functional outcomes (degeneracy), while the temporal energy accumulation $I_{Temporal}$ explains why individuals of different ages show the same kinetics (age independence). Both are rooted in the same fundamental property and together they define the *immune capacity* $\mathbf{I}$ as a set of invariants that govern the system's evolution.

### 4.3. A Falsifiable Prediction: Cross-Species Consistentency

The conservation law and the Immune Capacity $\mathbf{I}$ is derived from a symmetry principle that does not depend on the phylogenetic identity of the host organism. The structural symmetry is universal. Therefore, $\mathbf{I}$ should be consistent across species, provided the relevant parameters are matched.

So we predict if two subjects from different species are configured to possess identical values of the key systemic parameters, including the generalized masses $M$, the baseline driving forces $Q$, and the potential landscape $V$, then upon challenge with an identical immunogen, their subsequent immune responses must converge to statistically indistinguishable trajectories, after normalization by cellular turnover rates and metabolic scaling.

Preliminary evidence supporting this prediction comes from humanized mouse models. [19, 20]Transcriptomic profiling has revealed that humanized MHC mice exhibit significant similarity to human immune responses across both innate and adaptive immunity. Furthermore, selective expansion of human myeloid and NK cells in humanized mice promotes transcriptomic responses akin to those of human vaccinees. These observations suggest that when key systemic parameters are "humanized", the immune system's macroscopic dynamics converge toward human-like patterns, which is consistent with the conservation law's prediction that matched parameters yield matched outcomes across species.

## 5. The Properties and Biological Implications of the Immune Capacity

### 5.1. The Dimensionality of Immune Capacity

With the full Lagrangian and consistency constraints established, we now examine the dimensional structure of the framework.

The conformation coordinates $\mathbf{q_{ab}}, \mathbf{q_{ag}}$ describe the conformation of the antibody and the antigen, These are angular coordinates measured in radians, hence dimensionless. The appropriate inertial quantity is the moment of inertia and has dimension $[M][L]^2$

$\mathbf{r}$ describe the physical distance between the antibody and the antigen. The coordinates are positions in three-dimensional space, with dimension $[L]$. $M_r$ has dimension $[M]$.

Abundance coordinates $\xi$ describe the effective number of lymphocytes, It is dimensionless when normalized to a reference population size, The generalized mass $M_\xi$ has dimension $[M]$.

With these dimensions established, the structural invariant $I_{\text{Spatial}}$ has dimension $[M][L]^2[T]^{-1}$. The temporal invariant $I_{\text{Temporal}}$ has dimension of energy: $[M][L]^2[T]^{-2}$. The Immune Capacity **I** therefore combines a quantity with the dimension of action and a quantity with the dimension of energy.

### 5.2. Static Properties of the Immune Capacity

The structural invariant **I** quantifies the system's intrinsic resistance to conformational and compositional change.

A system with a large $I_{\text{Spatial}}$ has substantial momentum, it is undergoing rapid conformational changes, clonal expansions, or spatial rearrangements. Biologically, this corresponds to an active, ongoing immune response. A system with a small $I_{\text{Spatial}}$ is structurally stable, the conformations and clone sizes are changing slowly. This corresponds to homeostasis or a memory state where the system is poised but not actively responding.

Similarly, $I_{\text{Temporal}}$ leading to a consistency in energy variation. It characterizes the invariance of energy transfer during the immune system's evolution under a given generalized force $Q$. When this quantity is too low, the open immune system receives an insufficient net energy influx, resulting in a weak response that may fail to clear antigens efficiently and can lead to chronic infection.[21] If the value is too high, the immune reaction becomes excessively vigorous, potentially triggering pathological outcomes such as cytokine storms.

### 5.3. **The Reproducible Immune Dynamics under Antigen Perturbation**

The empirically observed reproducibility of immune responses across individuals finds its explanation in the statistical equivalence of the system's parameters. Different individuals of similar immune homeostasis possess potential landscapes $V_{self}$ that are statistically indistinguishable, because they are determined by conserved evolutionary constraints and shared physiological norms. The baseline generalized forces $Q_i^{net}$ are also comparable across individuals, as they reflect the organism's steady-state maintenance of the immune system including the tonic cytokine signaling, homeostatic proliferation signals, and metabolic inputs. When the pathogen with identical antigen conformation $\boldsymbol{q}_{ag}$, identical dosage $N_{ab}$ enters different hosts, the initial perturbation is therefore identical in magnitude and direction. The dynamical evolution governed by the extended Euler-Lagrange equation must yield the same macroscopic trajectory. This is the physical reason why different individuals, despite having different microscopic repertoires and different exposure histories, exhibit indistinguishable disease courses and antibody kinetics when challenged with the same antigen. The conservation law and its associated consistent evolution then ensure that this reproducibility holds not only for the initial response but throughout the entire course of the immune reaction.

### **5.4. The Role of Innate Immunity**

The influence of innate immunity and other systemic factors enters through the modulation of key parameters that define the system's initial conditions and boundary conditions. innate immunity modulates the effective initial values of several parameters. It reduces the effective antigenic impulse $p_{ag}$ by clearing pathogens through phagocytosis and complement. It enhances the coupling efficiency $G_k$ by promoting antigen presentation and co-stimulation through dendritic cells and macrophages. It modulates the generalized driving forces $Q_i^{in}$ by secreting cytokines that promote lymphocyte proliferation and survival.

A robust innate response reduces the antigenic invasion, resulting a smaller adaptive response; a compromised innate response results in a larger antigenic invasion, requiring a more vigorous (and potentially dysregulated) adaptive response.

## 6. The Measurement of Immune Capacity

### 6.1 Measurement of Key Parameters

The theoretical framework identifies the consistency constraint $\mathbf{I}$ as the physical embodiment of immune capacity, with its value determined by a set of experimentally accessible parameters. These parameters should be measured through background immune responses, such as intestinal immunity, enabling the framework to make predictions about the expected immune trajectory for any specified challenge prior to antigen encounter.

Antibody and antigen conformation coordinates $\boldsymbol{q}_{ab}, \boldsymbol{q}_{gb}$ can be determined through structural biology methods including X-ray crystallography and cryo-electron microscopy, or predicted with deep learning tools such as AlphaFold and RoseTTAFold. Spatial coordinates.[22] $\mathbf{r}$ are accessible through immunohistochemistry and tissue clearing combined with light-sheet microscopy. Cellular abundance coordinates $\boldsymbol{\xi}$, representing the effective number of lymphocytes in each clone, are routinely quantified by flow cytometry, mass cytometry (CyTOF), and single-cell RNA sequencing, with clonal identities established through immunoglobulin or T-cell receptor repertoire sequencing.

The dynamical parameters require dedicated experimental characterization. The generalized mass $M_{ab}, M_{ag}$ can be estimated from molecular dynamics simulations or from hydrogen-deuterium exchange mass spectrometry. $M_{\mathrm{r}}$ is simply the physical molecular mass, obtainable from its sequence or from size-exclusion chromatography. The $M_{\xi}$ associated with cellular abundance captures the inertial resistance of lymphocyte populations to changes in clone size. This quantity can be estimated by measuring immune cell cycle duration through nucleotide analog incorporation assays such as BrdU or EdU labeling, combined with flow cytometric tracking of division cycles. [23]

The immunogenicity coupling function $G$ can be assessed through in vitro stimulation assays that quantify the dose-response relationship between antigen exposure and immune activation. Standardized approaches include enzyme-linked immunospot (ELISPOT) assays for cytokine-secreting cells, intracellular cytokine staining followed by flow cytometry, and proliferation assays using dye dilution methods such as CFSE or CellTrace Violet. [24]The generalized driving force $Q^{in}$, monitored by multiplex immunoassays (e.g., Luminex or Olink) for key cytokines (IL-2, IL-4, IL-6, IFN-$\gamma$, TNF-$\alpha$), phospho-flow cytometry for STAT and NF-$\kappa$B pathway activities, and metabolomic profiling of lymphoid tissue supernatants. Integrating these measurements within the full dynamical framework enables the estimation of **I**, providing a quantitative basis for comparing immune capacity across individuals, disease states, and therapeutic interventions.

### 6.2 Quantifying Immune State via Repertoire-Derived Distance Metrics

Quantifying the immune state requires a measure that captures the system's position in the high-dimensional state space while remaining insensitive to the idiosyncratic stochastic details that distinguish one individual's repertoire from another's. Our laboratory previously introduced a computational strategy based on optimal transport theory that addresses this challenge by defining a distance metric between two immune repertoire states.[17] This immunological distance captures both its directional character and its magnitude. Whether the repertoire shifts toward clonally restricted populations dominated by a few expanded clones or maintains a balanced, polyclonal composition can be gauged by indicators such as the fraction of germinal-centre B cells, the prevalence of activated T-cell subsets, and the accrual of somatic hypermutations. The distance is formulated as the minimal cost required to transform one probability distribution of clonal abundances into another, with the cost function reflecting the biological expenditure associated with generating clones of different characteristics.

The practical utility of this approach has been demonstrated in both mouse models and human clinical samples. In mice subjected to different health conditions, the Wasserstein distance between experimental and reference repertoires discriminated across varying health strata in a manner consistent with known immunological phenotypes. More importantly, in a blinded clinical study, the framework distinguished paediatric febrile illness from Kawasaki disease without any training on labelled data. This training-free discrimination ability underscores the framework's robust predictive capacity and suggests that the Wasserstein distance captures biologically meaningful features of immune state that transcend the specific sequence identities of individual clones. The approach is fundamentally consistent with the symmetry-based conservation law developed in this work: the Wasserstein distance quantifies the macroscopic displacement of the immune system from homeostasis, while the consistency constraint **I** constrains the system's evolution, ensuring that different microscopic paths leading to the same displacement are dynamically equivalent. This repertoire-based distance method provides a surrogate measure of immune state without requiring direct measurement of velocities and generalised masses, making it a practical preliminary estimation tool that can be deployed using standard high-throughput sequencing data. As sequencing costs continue to decline and computational infrastructure improves, this optimal transport framework offers a scalable path toward quantitative immune state monitoring in clinical and research settings.

## 7. Conclusion and Discussion

In this work, we have addressed a fundamental paradox at the heart of modern immunology: how can adaptive immune responses exhibit reproducible macroscopic trajectories across different individuals, with molecular underpinnings are governed by stochastic processes including V(D)J recombination, somatic hypermutation, and clonal selection. We have shown that this apparent contradiction is resolved by recognizing that the immune system is governed by a continuous symmetry and its associated consistency constraint, derived from first principles through the Lagrangian formalism of analytical mechanics and Noether′s theorem. By constructing a dynamical description of the immune system in a high-dimensional state space spanned by generalized coordinates, and by identifying the invariance of the action under transformations, we have derived a consistency constraint $\mathbf{I}$ that acts as a global constraint on the system′s evolution. This consistency constraint, which we term the *Immune Capacity*, mandates that irrespective of the specific stochastic trajectory taken at the microscopic level, the system must converge to the same macroscopic endpoint for a given challenge, provided the initial conditions and boundary parameters are identical. The framework further explains why response kinetics depend only on elapsed time. Together, the symmetry provide a unified, first-principles explanation for the reproducibility of immune outcomes that has long been observed but never adequately explained.

The significance of this framework extends beyond resolving the reproducibility paradox. By partly transforming the phenomenological concept of immune homeostasis into a quantifiable physical invariant, our work establishes a foundational bridge between theoretical physics and immunology. The Immune Capacity $\mathbf{I}$ provides a rigorous operational definition for what clinicians and experimentalists have intuitively described as "strong" or "weak" immunity: a system with an appropriate $\mathbf{I}$ value responds proportionally to challenges, balancing activation and resolution. This quantitative framework unifies a diverse array of immunological phenomena as distinct dynamical states arising from the same underlying equations of motion, with each phenomenon corresponding to a specific configuration of the potential energy surface and the balance of driving and dissipative forces. The framework also yields a falsifiable prediction: if two organisms from different species are configured with identical consistent $\mathbf{I}$ and its associated parameters, then after appropriate allometric normalization of cellular turnover rates, their immune response trajectories must be statistically indistinguishable. This prediction, testable through humanized mouse models and standardized antigenic challenges, establishes the theory as a genuinely predictive rather than merely post-hoc descriptive framework.

The model employs a coarse-grained, mean-field approximation that averages over high-frequency regulatory fluctuations including circadian rhythms, hormonal cycles, and stochastic gene expression noise; these factors can be formally incorporated in future extensions as time-dependent modulations of the generalized driving forces $Q$. The sequential assumption of recognition before activation are valid at the single-cell level, represents an approximation at the population level where these processes overlap continuously; refining this to a fully continuous dynamical description would require integrating receptor engagement kinetics directly into the Euler-Lagrange equations. The parameters $G_k$ currently lack standardized measurement protocols, and future methodological work should focus on comparison of $\mathbf{I}$ values. More fundamentally, the framework as presented describes the adaptive immune system in isolation; integrating it with the neuro-endocrine axis, the microbiome, and metabolic regulation within a whole-organism perspective represents the next logical step toward a truly comprehensive theory of host defense.

Despite these limitations, by demonstrating that immune capacity is a consistency constraint arising from a symmetry principle, this work establishes a foundational framework for a predictive immunology grounded in physical law, an immunology in which the outcomes of host-pathogen encounters and therapeutic interventions can be understood, and ultimately predicted, from a set of first principles, offering a new paradigm for rational immunological design and quantitative assessment of immune health.

**Acknowledgements**

**Figure Resources**

Icons used in the figures are from Bioicons and Freepik.

**Funding**

This paper was supported by the National Key R&D Program of China (2023YFC2308200), National Science Foundation for Distinguished Young Scholars (82125037) and the National Natural Science Foundation of China(82230120).

**Author contributions:**

Yexing Chen: Conceptualization, Data curation, Formal analysis, Investigation, Methodology, Project administration, Software, Supervision, Validation, Visualization, Writing - original draft, Writing - review & editing

Peng Cao: Conceptualization, Project administration, Funding acquisition, Writing-review & editing

**Conflict of Interests**

The authors declare no conflict of interests.

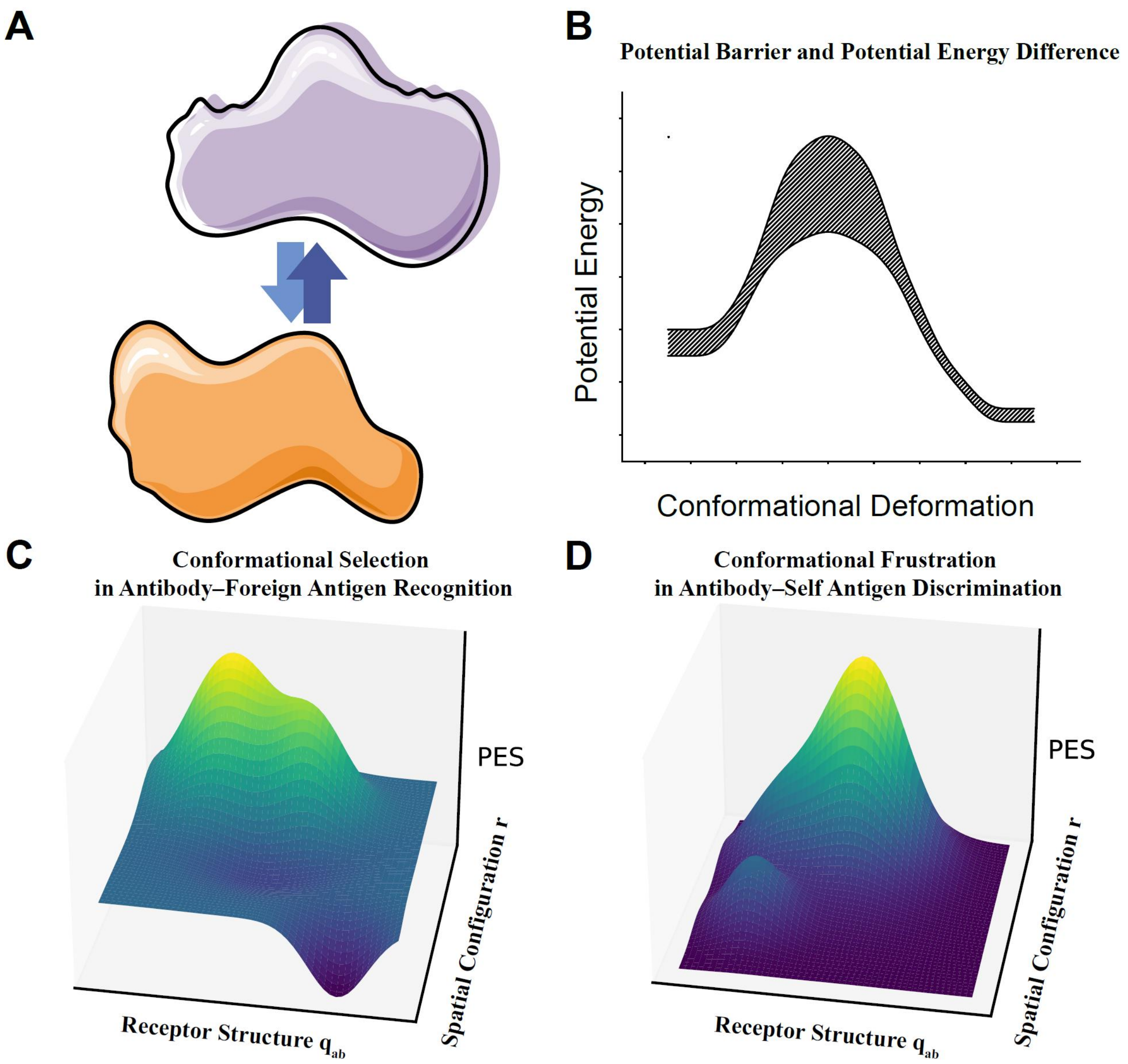

**Figure 1. The Potential Energy Surface of Immune Recognition. (A)** Foreign antigens collide with receptor proteins. Binding occurs based on the degree of structural match, as determined by the interaction energy. **(B)** Potential energy change during antigen-receptor binding. Within the active cellular subset, the energy barriers, post-binding steady-state energies, and released binding energies vary among different receptor conformations. This variance generates a distinct energy distribution profile, dictating the direction of subsequent adaptive system evolution. **(C)** The binding landscape for an entire immune receptor repertoire to a specific foreign antigen is represented as a PES. Only the the active cellular subset of receptor could achieve low-energy stable states. Upon encounter, the system evolves spontaneously toward the potential well. **(D)** PES for the binding of an immune receptor repertoire to a specific self-antigen. Immune tolerance is achieved by eliminating receptor clonotypes that would create deep potential wells for self-antigens, thereby preventing the initiation of autoimmune reactions.